\documentclass[reprint,twocolumn,showpacs,prb,aps,superscriptaddress]{revtex4-1}
\usepackage{graphicx}
\usepackage{amsbsy,amssymb,amsmath,bm,ulem,float}

\usepackage{color}

\graphicspath{{../fig/}}

\begin{document}

\title{Dynamics and morphology of dendritic flux avalanches in superconducting films}
\author{J. I. Vestg{\aa}rden}
\affiliation{Department of Physics, University of Oslo, P. O. Box
1048 Blindern, 0316 Oslo, Norway}
\author{D. V. Shantsev}
\affiliation{Department of Physics, University of Oslo, P. O. Box
1048 Blindern, 0316 Oslo, Norway}
\author{Y. M. Galperin}
\affiliation{Department of Physics, University of Oslo, P. O. Box
1048 Blindern, 0316 Oslo, Norway}
\affiliation{A. F.  Ioffe Physico-Technical
Institute of Russian Academy of Sciences, 194021 St. Petersburg,
Russia}
\author{T. H. Johansen}
\affiliation{Department of Physics, University of Oslo, P. O. Box
1048 Blindern, 0316 Oslo, Norway}

\begin{abstract}
We develop a fast numerical procedure for analysis of nonlinear
and nonlocal electrodynamics of type-II superconducting films in
transverse magnetic fields coupled with heat diffusion.  Using this
procedure we explore stability of such films with respect to dendritic
flux avalanches.  The calculated flux patterns are very close to
experimental magneto-optical images of MgB$_2$ and other
superconductors, where the avalanche sizes and their morphology change
dramatically with temperature.  Moreover, we find the values of a threshold
magnetic field which agrees with both experiments and linear stability
analysis. The simulations predict the temperature rise during an
avalanche, where for a short time $T \approx 1.5 T_c$, and a precursor
stage with large thermal fluctuations.
\end{abstract}

\maketitle

\section{Introduction}
The gradual penetration of magnetic flux in type-II superconductors
subjected to an increasing applied field or electrical current can be
interrupted by dramatic avalanches in the vortex matter.\cite{altshuler04}
 The mechanism
responsible for the avalanches is that an initial fluctuation reduces
locally the pinning of some vortices, which start to move, thus
creating dissipation followed by depinning of even more vortices. A
positive feedback loop is formed where a small perturbation can
escalate into a macroscopic thermomagnetic breakdown.\cite{mints81}

In thin film superconductors, the dynamics and morphology of these
avalanches is tantalizing, when at very high speeds they develop into
complex dendritic structures, which once formed remain robust against
changes in external conditions. When repeating identical experiments
one finds that the patterns are never the same although qualitative
features of the morphology, such as the degree of branching and
overall size of the structure, show systematic dependences on, e.g.,
temperature. Using magneto-optical imaging flux avalanches with these
characteristics have been observed in films of
Nb,~\cite{duran95,*welling04}
YBa$_2$Cu$_3$O$_{7-x}$,~\cite{brull92,*leiderer93, *bolz03}
MgB$_2$,~\cite{johansen02,albrecht05, *olsen07}
Nb$_3$Sn,~\cite{rudnev03} YNi$_2$B$_2$C,~\cite{wimbush04} and
NbN.~\cite{rudnev05,*yurchenko07} Investigations of onset conditions for the
avalanche activity have identified material dependent threshold values
in temperature,~\cite{johansen02} applied magnetic
field,~\cite{barkov03, yurchenko07} and transport
current,~\cite{bobyl02} as well as in sample size.\cite{denisov06}
Analytical modeling of the nucleation stage has explained many of
these thresholds using linear stability analysis.\cite{rakhmanov04,
aranson05, denisov05,denisov06}

Far from being understood is the development of the instability
from its nucleation stage to the fully developed dendritic pattern. 
Aranson et al.\cite{aranson05} explored the dynamics of the flux
avalanches as a numerical solution of Maxwell's equations with temperature 
dependent critical current density.  The dynamical process was 
governed by the interplay between an extremely nonlinear 
current-voltage relation, heat diffusion, 
and the nonlocal electrodynamics characteristic for thin superconducting  films. To
treat the nonlocal electrodynamics the authors used periodic
continuation of the sample taken as an infinite strip.  
This scheme should be a
good approximation inside the sample, although not necessarily close to the
edges. In fact, in thin films the magnetic field near the edges is
significantly enhanced~\cite{brandt93,*zeldov94} due to the flux expulsion.
Moreover, all experiments show that the instability
is always nucleated at an edge. Therefore, a careful account of the
electrodynamics close to the edges, including the  regions outside the
film, is expected to be crucially important.\cite{brandt95}  

In this work we study the formation and characteristics of dendritic flux avalanches using a numerical
scheme that takes into account the nonlocal electrodynamics 
both inside and outside a finite-sized superconducting film.
It is shown that our simulations largely reproduces experimental results obtained by 
magneto-optical imaging of dendritic avalanches in films of MgB$_2$, and 
furthermore gives detailed insight into not yet observed quantities such as local temperature rise
and electrical field.

The paper is organized as follows. 
Section~\ref{sec:model} presents the model and the equations describing the process.
The numerical scheme including the implementation of boundary conditions and 
thermomagnetic feedback is described in Sec.~\ref{num}.  The results for the
time-dependent distributions of magnetic flux and temperature are
presented and discussed in Sec.~\ref{sec:results}, while  Sec.~\ref{sec:conclusion} gives the conclusions.

\section{Model} 
\label{sec:model}

\begin{figure}[t!]
  \centerline{
    \includegraphics[width=0.9\columnwidth]{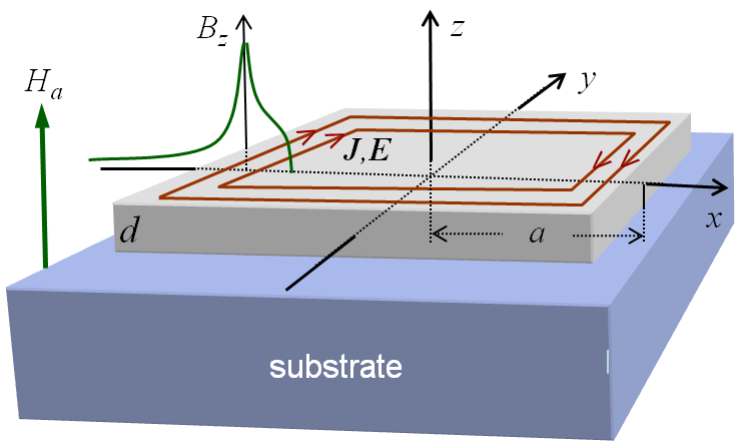}
  }
  \caption{(Color online) Schematic of the sample configuration.\label{fig1}}
\end{figure}

Consider a rectangular superconducting film zero-field cooled below the critical temperature, $T_c$, 
followed by a gradual increase in a perpendicular applied magnetic field.
The film is deposited on
a substrate, which in the process will be regarded as a sink for the dissipated heat.
Shown in Fig.~\ref{fig1} is a sketch of the overall configuration, including the relevant fields 
and currents.

The macroscopic behavior of  type-II superconductor films in a transverse
applied magnetic field, $H_a$, is well described by  quasi-static classical
electrodynamics.~\cite{brandt95-prl,brandt95}  Here the sharp depinning of vortices under flowing
current is represented by a highly nonlinear current-voltage relation
\begin{eqnarray}
  \label{power-law-EJ}
  \mathbf E &=&\rho (J)\mathbf J/d \, , \nonumber \\[2mm]
  \rho (J) & \equiv& \left\{
\begin{array}{lll}
 \rho_0 \left( J/J_c\right)^{n-1}, &  J \le J_c, & T \le T_c \, , \\
    \rho_0 \, , &  J > J_c, &T \le T_c \, , \\
    \rho_n\, , & & T > T_c\, . 
 \end{array}
\right.
\end{eqnarray}
Here $\mathbf E$ is the electric field, $\mathbf J$ is the sheet
current ($J \equiv |\mathbf{J}|$), $J_c$ the critical sheet current,
$n$ is the creep exponent, 
$\rho_0$ is a resistivity constant,  $\rho_n$ is the normal resitivity, and $T$ is temperature. 
It is assumed that the sample thickness, $d$, is so small that variations
in all relevant quantities across the thickness can be ignored.  
For $T\leq T_c$ the temperature
dependence of the critical current and  flux creep
exponent~\cite{denisov06}  are taken as
\begin{equation}
J_c = J_{c0}(1-T/T_c)  \ \  {\rm and} \  \quad n-1 = n_0 T_c/T \, ,
\end{equation}
where  $J_{c0}$ and $n_0$ are constants.

The distribution of temperature is described by  the heat diffusion equation
\begin{equation}
   \label{dynamics-T}
   dc \, \dot T  =  d\nabla\cdot (\kappa \nabla T)  - h(T-T_0)+
   \mathbf J\cdot \mathbf E  \,  , 
\end{equation}
where $\kappa$ is the thermal conductivity of the superconductor, $c$ is
its specific heat, $T_0$ is the substrate
temperature, taken to be constant, and $h$ is the coefficient of heat transfer between the
film and the substrate. The  $\kappa, c$ and $h$ are all
assumed to be proportional to $T^3$, whereas a
relatively weak temperature dependences of $\rho_0$ and $\rho_n$ are
neglected.~\cite{schneider01,denisov06}  

Following Ref.~\onlinecite{brandt95} we define the local magnetization, 
$g=g(\mathbf r)$, as 
\begin{equation}
  \label{cur1}
  \nabla g\times \mathbf{z} =\nabla \times (g\mathbf{z}) =  \mathbf{J} \ ,
\end{equation}
where $\mathbf r\equiv  (x,y)$ is a 2D vector in the film plane, and
$\mathbf{z}$ is the unit vector in the perpendicular direction. 
Outside the sample there are no currents, and we set $g=0$ by definition.  
The Biot-Savart law can then be written as
\begin{equation}\label{biot-savart}
  \frac{B_z(\mathbf r)}{\mu_0} -H_a=\hat{Q}g \equiv \int d^2r'\,
  Q(\mathbf{r}-\mathbf{r}',z) g(\mathbf{r}') \, ,
\end{equation}
where the integral is calculated over the whole plane. The kernel
$Q(\mathbf{r})$ should be calculated as a limit at $z\to 0$ of  the
expression
\begin{equation} \label{biot-savart1}
  Q(\mathbf{r},z)=\frac{1}{4\pi}\frac{2z^2-r^2}{(z^2+r^2)^{5/2}}\, , \ r
  \equiv |\mathbf{r}|\, .  
\end{equation}
Here reqularization is needed to avoid formal divergence of the
r.h.s. of Eq.~\eqref{biot-savart} at $z=0$, $\mathbf{r}=\mathbf{r}'$. 
The Fourier transform of $\lim_{z\to 0}Q(\mathbf{r},z)$ is equal to
$k/2$.\cite{roth89} 
Therefore, from the convolution theorem it follows that the inverse
operator $\hat{Q}^{-1}$ acting on some function $\varphi (\mathbf{r})$
can be expressed as  
\begin{equation} \label{hatQ}
\hat{Q}^{-1} \varphi(\mathbf{r}) =2\mathcal{F}^{-1} \left(k^{-1}
  \mathcal{F}[\varphi(\mathbf{r})] \right)\, . 
\end{equation}
Here ${\mathcal F} [\varphi (\mathbf{r})]$ and $ {\mathcal
  F}^{-1}[\varphi (\mathbf{k})]$  are Fourier and inverse Fourier
transform, respectively,  and $k \equiv |\mathbf k|$.

Inverting Eq.~(\ref{biot-savart}) one arrives at the equation for the
time evolution of the local magnetization, 
\begin{equation}\label{se1}
\dot{g}(\mathbf{r},t)=2\mathcal{F}^{-1}\left( k^{-1} \mathcal{F}
\left[\mu_0^{-1} \dot{B}_z (\mathbf{r},t)-\dot{H}_a(t)\right]\right) \, .
\end{equation}
Equations~\eqref{dynamics-T}, \eqref{cur1}  
and \eqref{se1} therefore determine the dynamics of $g(\mathbf{r},t)$, 
 $T(\mathbf{r},t)$, etc. 
To solve these equations numerically we proceed from the continuous to
a discrete formulation.

\section{Numerical approach} \label{num}

To allow use of the fast Fourier transform (FFT) we consider a rectangular 
area of size $2L_ x \times 2L_y$ containing the sample
plus a substantial part of its surrounding area. 
A key point is to select  proper values for $L_x$ and $L_y$
relative to the sample size,  $2a \times 2b$.
By including too little area outside the sample one clips away the slowly 
decaying tail of the stray fields, leading to decreased accuracy at large scales, and 
major deviations from the correct physical behavior.~\cite{brandt95}
On the other hand, including too much of the outside area 
keeping the same number of the grid points  tends to decrease 
the accuracy at small scales, where actually
the most interesting features of the dendritic avalanches appear. 
This blurring can be compensated by using a finer spatial grid, 
at the cost of a rapidly increasing computation time.

A careful test of our numerical scheme was done by comparing the calculations with
the exact solution for the Bean critical state in an infinitely long strip.~\cite{brandt93,*zeldov94}
It is found  that already with $L_x/a \gtrsim 1.3$ the calculated  results are correct
within a few percent, and are essentially indistinguishable from the exact solution in graphic comparisons.

In the FFT-based calculations
the rectangle $2L_ x \times 2L_y$  is discretized as a $N_x\times N_y$ equidistant grid, 
and used as unit cell in an infinite superlattice. The Fourier wave vectors $k_{x,y}$ are then discrete, 
$k_{x,y}= \pi q_{x,y}/L_{x,y}$, where  $q_{x,y}$ are integers.  
The Brillouin zone is chosen as $|q_{x,y}| \le N_{x,y}/2$, which ensures 
$g(\mathbf{r},t)$, $T(\mathbf{r},t)$, etc. to be real valued.

The calculation of the temporal evolution is based on  a 
discrete integration forward in time
\footnote{The discrete time integration is explained using Euler's method, 
but the actual implementation uses the Runge-Kutta method.}
of the local magnetization
\begin{equation}
  g(\mathbf r,t+\Delta t) \approx g(\mathbf r,t)+\Delta t~\dot g(\mathbf r,t) \, ,
  \label{dynamics-g}
\end{equation}
starting from $g(\mathbf r,0)=0$. 
Once $g(\mathbf r,t)$ is known at time $t$, we proceed one time step by 
determining $\dot g(\mathbf r,t)$. The $\dot{g}(\mathbf{r},t)$ 
can be calculated from Eq.~\eqref{se1}, provided $\dot{B}_z$ 
is known \textit{everywhere} within the unit cell.
For this, we have to find self-consistent solutions 
for $\dot{g}$ and  $\dot{B}_z$ given the function $g$.

For the area inside the superconductor  the material law,  
Eq.~\eqref{power-law-EJ}, applies 
and together with the Faraday law, $\dot{B_z}=- (\nabla \times \mathbf{E})_z$, 
it follows that 
\begin{equation}
  \label{se2}
  \dot B_z = \nabla \cdot (\rho\nabla g)/d\, .
\end{equation}
The gradient $\nabla g(\mathbf r,t)$ is readily calculated, and since 
the result allows finding $\mathbf J(\mathbf r,t)$, from Eq.~\eqref{cur1},
also $\rho(\mathbf r,t)$ is determined from Eq.~\eqref{power-law-EJ}.
The difficult point is that $\dot{g}$ depends on the distribution of $\dot{B}_z$ in the
whole unit cell. The task is to find the $\dot{B}_z$ outside the sample 
which leads to $\dot g=0$ outside. This cannot be calculated directly 
since there is a nonlocal relation between
$\dot{B}_z$ and  $\dot{g}$. Instead we use an iterative procedure. 

Let us label the iterations by a superscript $(i)$. At the first step, $i =1$, 
we calculate $\dot{B}_z$ \textit{inside}
the superconductor from Eq.~\eqref{se2}.  Then  an initial
guess is made for the time derivative, $\dot{B}_z^{(1)}$, 
\textit{outside} the sample. From Eq.~\eqref{se1} we now compute the time derivative 
$\dot{g}^{(1)}$.  In general, this $\dot{g}^{(1)}$ does not vanish
outside the superconductor.
To correct for this, a new and improved  $\dot{B}_z$ is chosen as 
\begin{equation}
  \dot{B}_z^{(i+1)} = \dot{B}_z^{(i)}  -\mu_0\hat{Q}\hat{O} \dot{g}^{(i)} +C^{(i)} 
  ,
\end{equation}
where the projection operator $\hat O$ vanishes inside the
superconductor and equals to 1 outside it.  The constant $C^{(i)}$ is
determined by the flux conservation,
\begin{equation}
  \int d^2r\, [\dot B_z^{(i+1)}(\mathbf r,t)-\mu_0 \dot H_a]=0  
  .
\end{equation}
The procedure is stopped
after $s$ iterations when the values of $\dot{g}$ outside the
superconductor becomes sufficiently small.  The final
distribution, $\dot{g}^{(s)}(\mathbf{r})$, is taken as the ``true''
$\dot{g}(\mathbf{r},t)$, and substituted into 
Eq.~\eqref{dynamics-g} in order to advance in time.

A good choice for the initial state of the iteration at time $t$ is
$\dot{B}_z^{(1)}(t)=\dot {B}_z^{(s)}(t-\Delta t)$, i.e., each
iteration starts from the final distributions achieved during the
previous iteration.  Normally, 
$s=5$ iterations is sufficient to give good results.

\begin{figure}[t]
  \begin{center}
    \includegraphics[width=\columnwidth]{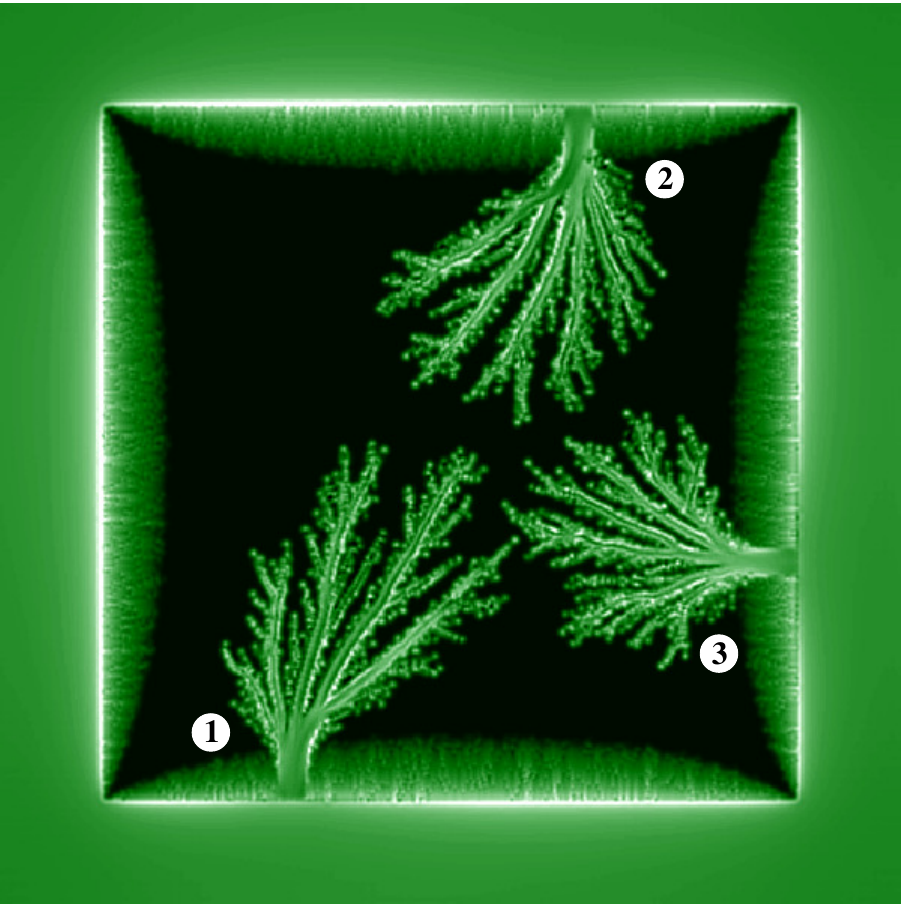} 
  \end{center}
  \caption{
    (Color online)
    Calculated distribution of $B_z$ at an applied field  $H_a=0.18 J_{c0}$, and  
    substrate temperature $T_0=T_c/4$. 
    The image brightness represents the magnitude of $B_z$.
    The sample contour appears as a bright
    rim of enhanced field, and the black central area is the flux-free
    Meissner state region. 
}
    \label{fig:b}
\end{figure}

\section{Results and discussion} \label{sec:results}

Numerical simulations were performed for samples shaped 
as a square of side $2a$ and with an outside area
corresponding to  $L_x = L_y =1.3a$. 
The total area is discretized on a 512$\times$512 equidistant grid. 
Quenched disorder is included in the model by a 
10\% reduction of $J_{c0}$ at randomly selected
5\% of the grid points.  The simulated flux penetration process starts at zero applied
field with no flux trapped in the sample, which has a uniform temperature
$T_0$.

\begin{figure}[t]
  \begin{center}
    \includegraphics[width=\columnwidth]{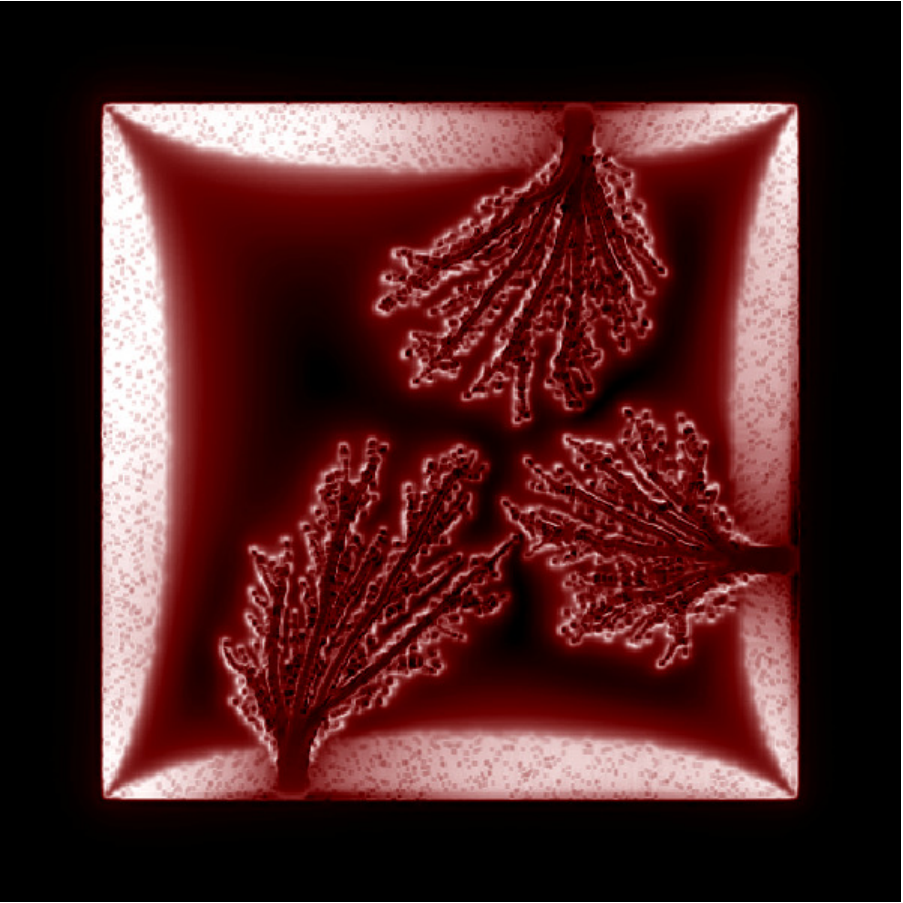} 
  \end{center}
  \caption{
    (Color online)
    Map of the sheet current, $J$,
    corresponding the image in Fig.~\ref{fig:b}.
    The brightness represents $J$, where 
    black means $J=0$. 
  }
    \label{fig:j}
\end{figure}

\begin{figure}[t]
  \begin{center}
    \includegraphics[width=\columnwidth]{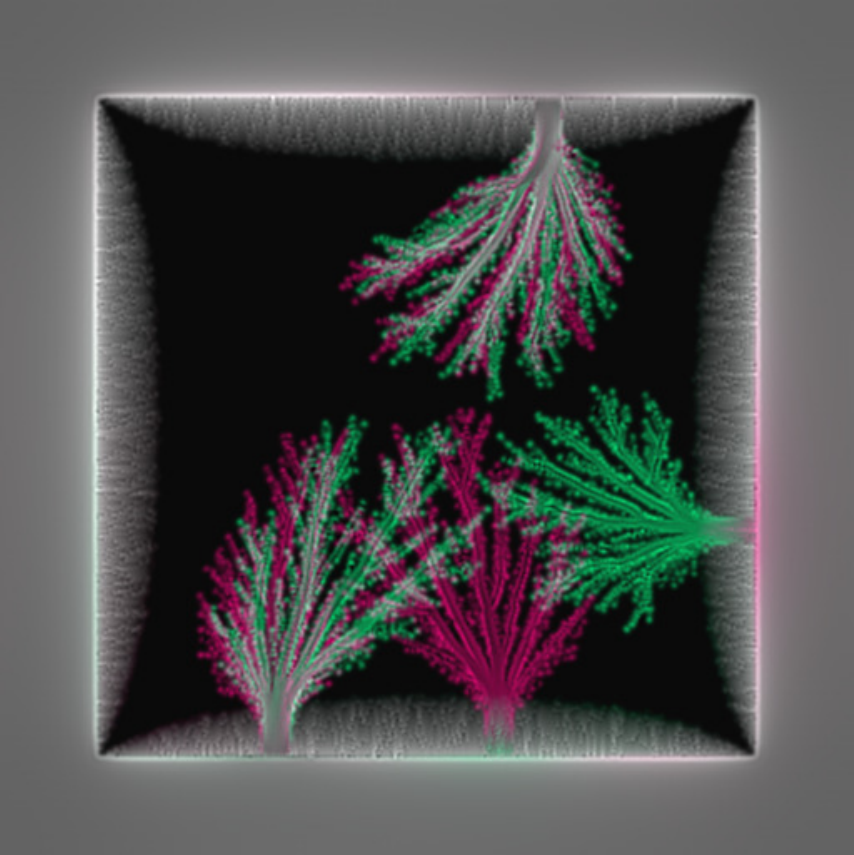} 
  \end{center}
  \caption{
    (Color online)
    Color-coded overlay of two separate runs with same quenched disorder but    
    with different microscopic fluctuations. The pixels in gray-scale
    represent overlapping results.  The parameters are the same as in the 
    caption of Fig.~\ref{fig:b}.
  }
    \label{fig:z}
\end{figure}

Calculations were performed at $T_0=T_c/4$ using material parameters corresponding to 
a typical MgB$_2$ film,\cite{schneider01, denisov06}  
$\rho_n$=7~$\mu\Omega$cm, $\kappa = 0.17$~kW/Km$\times (T/T_c)^3$ and
$c = 35 $ kJ/Km$^3\times (T/T_c) ^3$, where $\rho_n$ is the normal
resistivity at $T_c=39~$K, $J_{c0}=50$~kA/m, $\rho_0=\rho_n$,
$d=0.5~\mu$m, $a=2.2~$mm, and $h=220$~kW/Km$^2\times
(T/T_c)^3$.
We choose $n_0=19$ and limit the creep exponent to 
$n(T)\leq n_\text{max}=59$.
The field was ramped from $H_a=0$ at a constant rate, $\dot
H_a=10^{-5}J_{c0}\rho_n/ad\mu_0$.

Figure~\ref{fig:b}
shows the $B_z$-distribution at $\mu_0H_a=0.18 \mu_0J_{c0}=11$~mT, 
where three large dendritic structures have already been formed. 
The numerical labels indicate
the order in which they appeared during the field ramp. The first
event took place at the threshold applied field,  $\mu_0H_{\rm th} = 0.145 \mu_0J_{c0}=9.1$~mT,
which is in excellent
agreement with measurements on MgB$_2$ films just below 10~K$\approx T_c/4$.
At lower fields, the flux penetration was gradual and smooth,
just as seen on the left edge of the sample, where the
characteristic ``pillow effect'' for films in the critical state is
very well reproduced.~\footnote{Note a slight corrugation in this smooth
pattern, which originates from the slightly nonuniform $J_{c0}$, a
detail commonly seen in magneto-optical images of real samples.}

The dendritic avalanches all nucleate at the edges, and one by one
they quickly develop into a branching structure that extends far
beyond the critical-state front and deep into the Meissner state
area. The trees are seen to have a morphology that strongly
resembles the flux structures observed experimentally in many
superconducting films.~\cite{leiderer93,
bolz03,duran95,welling04,johansen02,albrecht05,
olsen07,rudnev03,wimbush04,yurchenko07} 
The simulations also reproduce the experimental finding that
once a flux tree is formed, the entire dendritic structure remains
unchanged as $H_a$ continues to increase. The supplementary material
\footnote{
See Supplemental Material at URL for VIDEO clips 
showing the development of $B_z$ with time.}
includes a VIDEO clip of the dynamical process, and shows striking
resemblance with magneto-optical observations of the phenomenon.

Figure~\ref{fig:j} shows the sheet current magnitude,
$J$, corresponding to the flux distribution in Fig.~\ref{fig:b}. 
From
this map it is clear that the dendrites completely interrupt
the current flow in the critical state, and redirect it around the
perimeter of the branching structure. This vast perturbation of the
current has been demonstrated experimentally earlier using inversion of
magneto-optical images.~\cite{laviano04, *olsen06}
Note that the critical state region contains dark pixels 
which are  the randomly distributed sites of reduced $J_{c0}$.

To investigate reproducibility in the pattern formation, microscopic
fluctuations were introduced by randomly alternating between right-
and left-derivatives in the discrete differentiation. Due to the
nonlinear form of Eq.~\eqref{power-law-EJ} this procedure gives large
local variations in the electrical field.  Figure~\ref{fig:z}
shows an overlay of two simulation runs with different realizations of
the microscopic fluctuations while keeping the same quenched disorder
in $J_{c0}$. The two resulting images were colored so that adding them
gives shades of gray where both coincide in pixel values. Clearly, the
two runs gave different results as far as the dendritic pattern is
concerned. Both produced three branching structures, where two are rooted
at the same place and the third is at a different location. 
\footnote{The two roots overlap because clustering of the quenched disorder 
facilitate nucleation of the thermomagnetic instability.} 
Even for
those with overlap, there are parts of the structure that differ
considerably, especially in the finer branches. In contrast, both the
critical state and the Meissner state regions are essentially
identical in the two runs. Note the color at the edge of 
the right hand side near the root 
of the green dendrite, which reflects that the growth of the flux
structure drains the external field near the root.
 Moreover, the root
of all the trees are not far from the middle of the sides. Both
features are in full accordance with experiments.

\begin{figure}[t]
  \centerline{
    \includegraphics[width=\columnwidth]{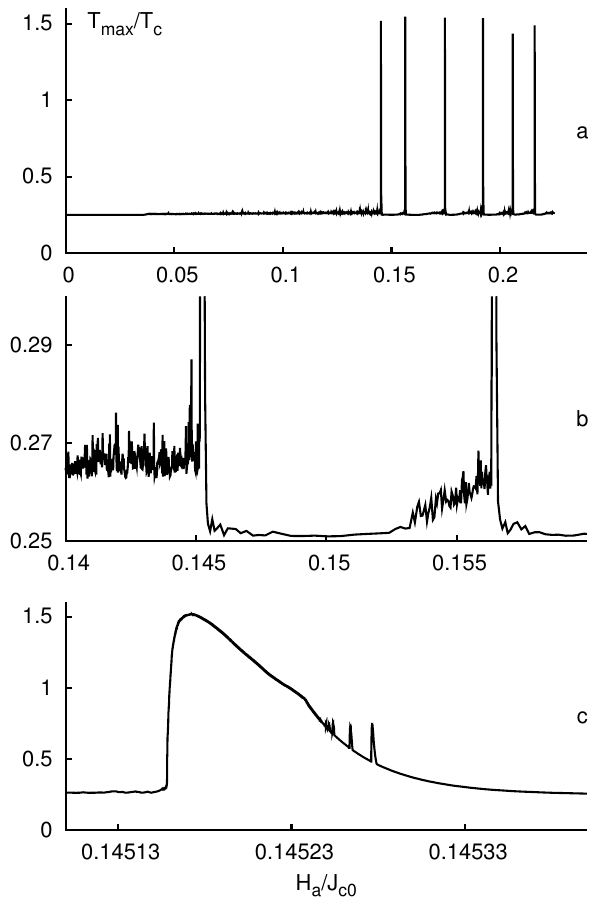}
  }
  \caption{
    Maximum temperature
    in the superconductor during an ascending field ramp at $T_0=T_c/4$.
    The  panels (a)-(c) are successive magnifications of the first avalanche event.
    \label{fig:temp}
  }
\end{figure}

Each dendritic avalanche is accompanied by a large local increase in
temperature. Shown in Fig.~\ref{fig:temp}a is a plot of the maximum
temperature in the film during a field ramp with substrate kept at
$T_0 = T_c/4$.  
The spikes in the temperature rise as high as $ 1.5 T_c$. 
The maximum temperature is found in the root region of the avalanche.
The heating above $T_c$ is an interesting prediction; to our
knowledge, the temperature of propagating avalanches has not been
observed experimentally. At the same time, the result is consistent with the
measured heating of uniform flux jumps in Nb foils \cite{prozorov06}
and the magnetic field-induced damage in a YBa$_2$Cu$_3$O$_{7-x}$ film during
dendritic growth.\cite{brull92}

\begin{figure}[t]
   \centerline{
     \includegraphics[width=\columnwidth]{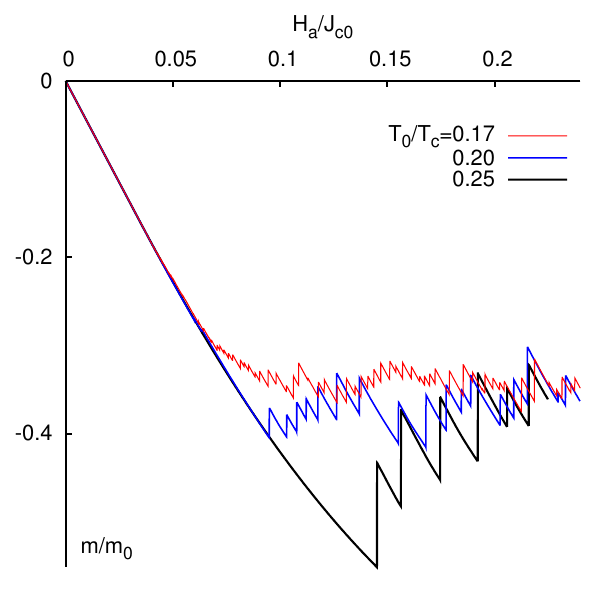}
   }         
  \caption{
    (Color online)
    Magnetic moment in units of $m_0=a^3J_{c0}$ as function of
    increasing field obtained by simulations at three different temperatures, $T_0$.
    Each jump in the curves represents a flux avalanche.
    \label{fig:moment}
  }
\end{figure}
\begin{figure*}[t]
 \centerline{
   \includegraphics[width=\textwidth]{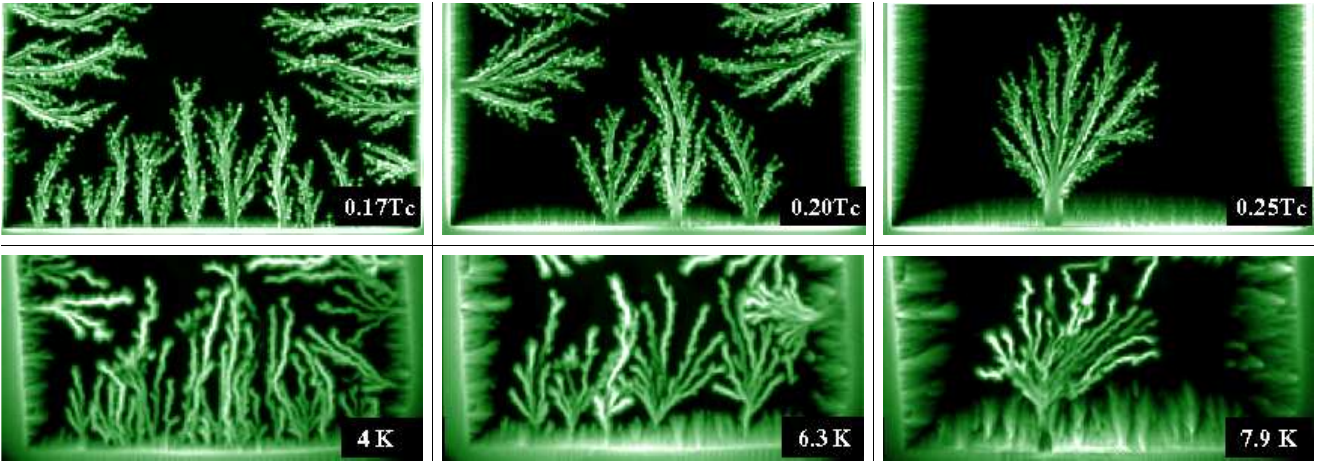}
   }  
  \caption{
    (Color online)
    Temperature variation in the morphology of flux dendrites.
    Top panels show simulated  results for $B_z$  and
    bottom panels show magneto-optical images of a MgB$_2$ film.
    \label{fig:temp3}
  }
\end{figure*}

The first avalanche in Fig.~\ref{fig:temp}a appears 
at  $H_{\rm th} = 0.145 J_{c0}$.
Since the chosen disorder is rather weak and the
ramp rate is high,  the heat diffusion to the substrate 
is expectedly a more important stabilizing factor than lateral heat diffusion, 
the theoretically predicted threshold field is~\cite{denisov06}
\begin{equation}
  \label{Hth}
  H_\text{th} = \frac{J_{c}'}{\pi}\tanh^{-1}\left(\frac{T_ch}{naJ_{c0}\mu_0\dot H_a}\right)\, .
\end{equation}
At $T=T_c/4$ and with $n=59$ this  gives $H_\text{th}=0.15J_{c0}$, in excellent
agreement with the present simulation.  Here, $J_{c}'=0.6J_{c0}$ is the
effective critical current, which is lower than $J_c$ due to flux
creep. At the same time, the adiabatic threshold field
\cite{denisov05} is much smaller than $H_\text{th}$, which means 
that the heat diffusion and  heat transfer to the substrate prevent
avalanches. However, during short time intervals cooling
is not always effective, and the temperature experiences large
fluctuations. The fluctuations are particularly large as $H_a$
approaches triggering of an avalanche, see Fig.~\ref{fig:temp}b.  In
these intervals both heat absorption and lateral heat diffusion play
important roles in stabilizing the superconductor.  A close-up view of
the maximum temperature during the first avalanche at $T_0 = T_c/4$ is
shown in Fig.~\ref{fig:temp}c.  First, the temperature rapidly
increases, and then decays much slower.  The duration of the avalanche
is $0.18~\mu$s. Since the length is $2.5~$mm, the average  propagation velocity 
is of order $14$~km/s. This numerical value is reasonable compared to 
previous measurements, where the flux dendrites were triggered by a laser
pulse in YBaCuO films.\cite{leiderer93,bolz03} The maximum electric
field in the superconductor during the avalanche is also high, found from 
the simulations to be approximately $5$~kV/m.

The abrupt redirection of the current implies that the magnetic
moment of the sample makes a jump and becomes smaller.
Figure~\ref{fig:moment} shows the moment as function of the increasing
applied field. Each vertical step corresponds to a flux avalanche.
The lower curve, obtained for $T_0 = T_c/4$, shows jumps with typical
size of $0.1m_0$ with a slight dispersion, which is due to variations
both in shape and location of the avalanches. More pronounced is the
variation in jump size with temperature. As $T_0$ gets lower the
jump size becomes smaller, and the events more frequent. In the
graphs for $T_0/T_c = 0.20$ and 0.17, the jump size reduces to $0.03
m_0$ and $0.01 m_0$, and jumps appear on average with field intervals
of $\Delta H_a/J_{c0} = 0.01$ and $0.002$, respectively. In real samples 
a similar 
temperature variation of jumps in the $m$-$H$ curves was observed by
magnetometry.~\cite{zhao02,rudnev03,prozorov06,colauto08}

It has been reported~\cite{johansen02} that the morphology of flux
avalanches is strongly temperature dependent.  This is illustrated in
the bottom panel of Fig.~\ref{fig:temp3} showing three magneto-optical
images of a 0.4~$\mu$m thick MgB$_2$ square film at $T_0=$4~K, 6.3~K
and 7.9~K. The images show a crossover from many long fingers at 4~K
to medium sized dendrites at 6.3~K, to a single highly branched
structure at 7.9 K.  The simulation results shown in the top panels reproduce this
result and show exactly the same trend as the experiments.  At the
lowest temperature, $0.17T_c$, there are many finger-like avalanches.
At the middle temperature $0.2T_c$ there are fewer avalanches, with
typically three to four branches each. At the highest temperature
$0.25T_c$ there is just one big avalanche, with seven main branches.

\section{Conclusion}
\label{sec:conclusion}
In conclusion, we have developed and demonstrated the use of a fast numerical scheme for simulation
of nonlinear and nonlocal transverse magnetic dynamics of type-II
superconducting films under realistic boundary conditions.  Our 
simulations of thermomagnetic flux avalanches qualitatively and quantitatively
reproduces numerous experimentally observed features: the fast flux
dynamics, morphology of the flux patterns, enhanced branching at
higher temperatures, irreproducibility of the exact flux patterns,
preferred locations for nucleation, and the existence of a threshold field.  The scheme
allows determination of key characteristics of the process such as maximal
values of the temperature and electric field 
as well as typical propagation velocity. 

\acknowledgments The work was supported financially by the Norwegian
Research Council. We are thankful to M.~Baziljevich for helpful
discussions.

%


\end{document}